\documentclass[aps,prl,showpacs,groupaddress,twocolumn,superscriptaddress,floatfix]{revtex4-1}
\usepackage{graphicx}
\usepackage{dcolumn,float}
\usepackage{amsmath,amssymb}
\usepackage{subfigure}

\newcommand{\be}{\begin{eqnarray}}

\newcommand{\ee}{\end{eqnarray}}

\newcommand{\qq}{\begin{eqnarray}}
\newcommand{\qqq}{\end{eqnarray}}

\newcommand{\beg}{\begin{equation}}
\newcommand{\en}{\end{equation}}

\begin{document}

\title{Enhancement of the critical temperature in iron-pnictide superconductors by finite size effects}
\author{M. A. N. Ara\'ujo}

\affiliation{Departamento de F\'{\i}sica,  Universidade de \'Evora, P-7000-671, \'Evora, Portugal}
\affiliation{CFIF, Instituto Superior T\'ecnico, 
Universidade T\'ecnica de Lisboa,  Av. Rovisco Pais, 1049-001 Lisboa, Portugal}

\author{Antonio M. Garc\'{\i}a-Garc\'{\i}a}
\author{P. D. Sacramento}
\affiliation{CFIF, Instituto Superior T\'ecnico, 
Universidade T\'ecnica de Lisboa, Av. Rovisco Pais, 1049-001 Lisboa, Portugal}

\begin{abstract}
Recent experiments have shown that, in agreement with previous theoretical predictions, superconductivity in metallic nanostructures can be enhanced with respect to the bulk  ($L \to \infty$) limit. 
Motivated by these results we study finite size effects (FSE) in an iron-pnictide superconductor.  
For realistic values of the bulk critical temperature $T_c^{\rm bulk} \sim 20-50$K, we find that, in the nanoscale region $L \sim 10$ nm, $T_c(L)$
has a complicated oscillating pattern as a function of the system size L. A substantial enhancement
of $T_c$ with respect to the bulk limit is observed for different boundary conditions, geometries and
two microscopic models of superconductivity. Thermal 
fluctuations, which break long
range order, are still small in this region. Finally we show that the differential conductance, an
experimental observable, is also very sensitive to FSE.
\end{abstract}

\pacs{74.70.Xa,74.78.Na,74.20.Fg,74.78.-w}

\maketitle

\newcommand{\bb}{\boldsymbol{\beta}}
\newcommand{\ba}{\boldsymbol{\alpha}}

Recent technological developments \cite{chi1,tin1} have opened the possibility for the synthesis and characterization of high quality nano-structures, a recurrent problem in previous \cite{oldexp} experimental studies of superconductivity in the nanoscale region. 
 As an example, it is now possible \cite{natm} to track experimentally the evolution of superconductivity in single, clean and isolated metallic nanoparticles of typical size $L > 5$ nm by using low temperature scanning tunnelling microscopy/spectroscopy.\\ These advances have revived the interest in nanoscale superconductors.
Theoretical studies on this problem were pioneered by Anderson \cite{ander2} more than fifty years ago. At zero temperature the exact Richardson's equations \cite{richardson} provides a satisfactory account of quantum fluctuations and 
other deviations from mean-field behaviour in nanosuperconductors. At finite temperature the role of thermal fluctuations is well described by the path integral formalism \cite{scala}.
In the limit of negligible disorder, FSE in conventional superconductors have been studied for different systems: an harmonic oscillator \cite{heiselberg}, rectangular nanofilms \cite{peeters,blat,parmenter}, a nanowire \cite{peeters}, a sphere \cite{sphere}, and a chaotic grain \cite{usprl,leboeuf}.  
The main conclusions of these studies are that: a) FSE can enhance or suppress $T_c$ substantially provided that the superconducting coherence length $\xi$ is not much smaller than $L$, b) FSE are stronger the smaller and more symmetric the grain is, c) thermal fluctuations are controlled by the parameter $\gamma = \beta\sqrt{\delta/T_c}$ \cite{scala} where $\delta$ is the mean level spacing around the Fermi energy and $\beta \approx 1/2$. For $\gamma \ll 1$ long-range order still holds as only a narrow region $\sim \gamma T_c$ around $T_c$ is affected by thermal fluctuations. These conditions set the minimum $L$ for which enhancement of $T_c$ is observed.\\
A natural question to ask, especially after the recent discovery of iron pnictides \cite{iron}, is whether these results are of relevance for high $T_c$ superconductors. 
To the best of our knowledge not much is known about FSE in
high $T_c$ materials.
Cuprate nanowires \cite{nanohightc} and carbon nanotubes \cite{paola} have been recently
investigated experimentally for sizes where thermal and quantum fluctuations are dominant. We are not aware of theoretical studies of FSE in high $T_c$ superconductors (for a recent study on multiband superconductors see \cite{bianconi}). 
This paper is a first step in this direction.\\ 
We investigate superconductivity in clean, rectangular nanofilms of iron pnictides. All our calculations are in two dimensional geometries as it is believed that this is the effective dimensionality relevant for superconductivity in these materials. Iron-pnictides are more suitable than cuprates because, unlike cuprates, superconductivity in iron pnictides seems to be well described by 
a mean field approach based on Fermi liquid theory. This is a clear advantage as 
the ideas and techniques introduced in conventional nanosuperconductors are still applicable. Moreover FSE are easier to detect in iron-pnictides as its  
$\xi$ is larger than that of cuprates.\\
\begin{figure}
\includegraphics[height=0.5\columnwidth,clip]{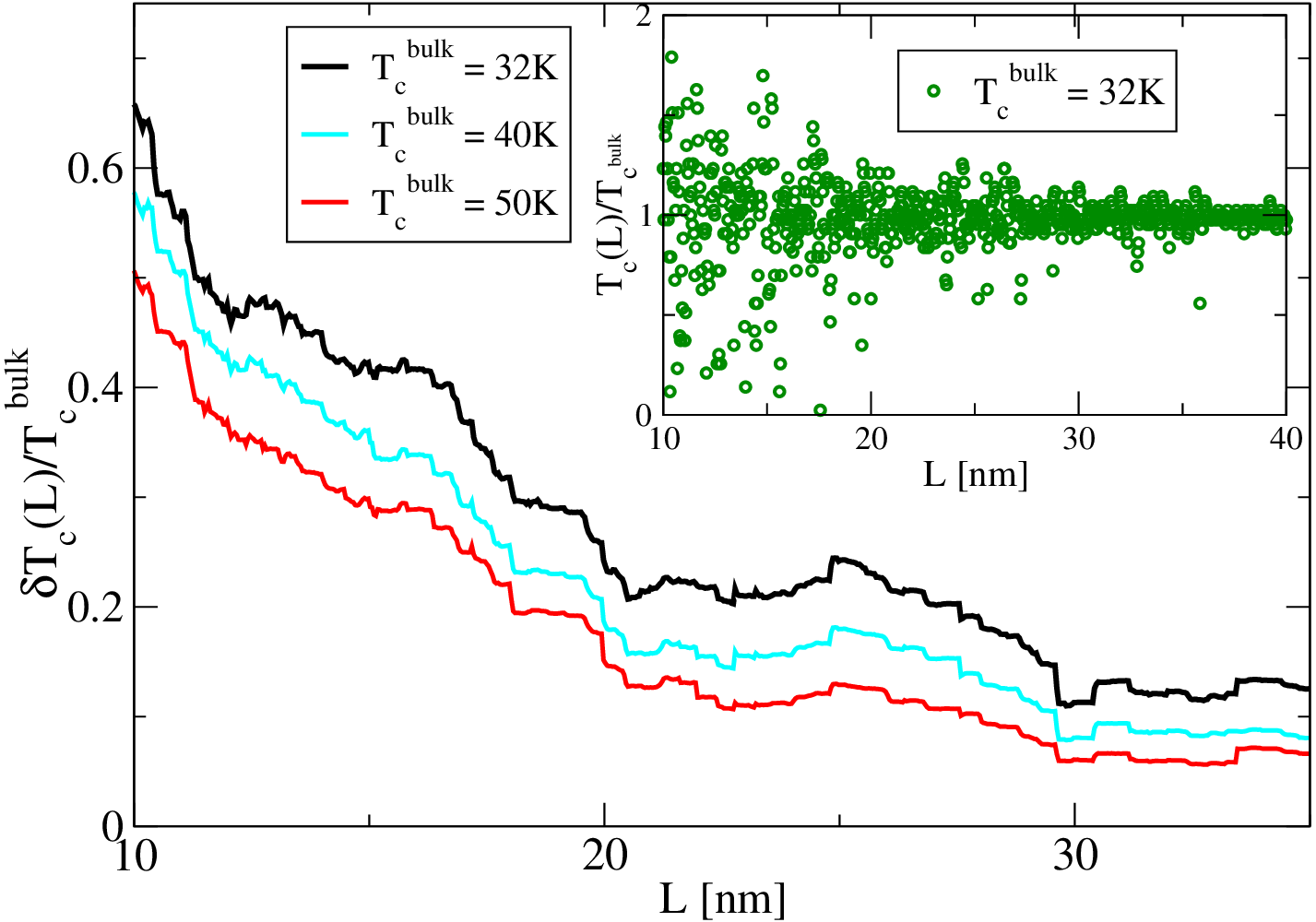}
\includegraphics[height=0.5\columnwidth,clip]{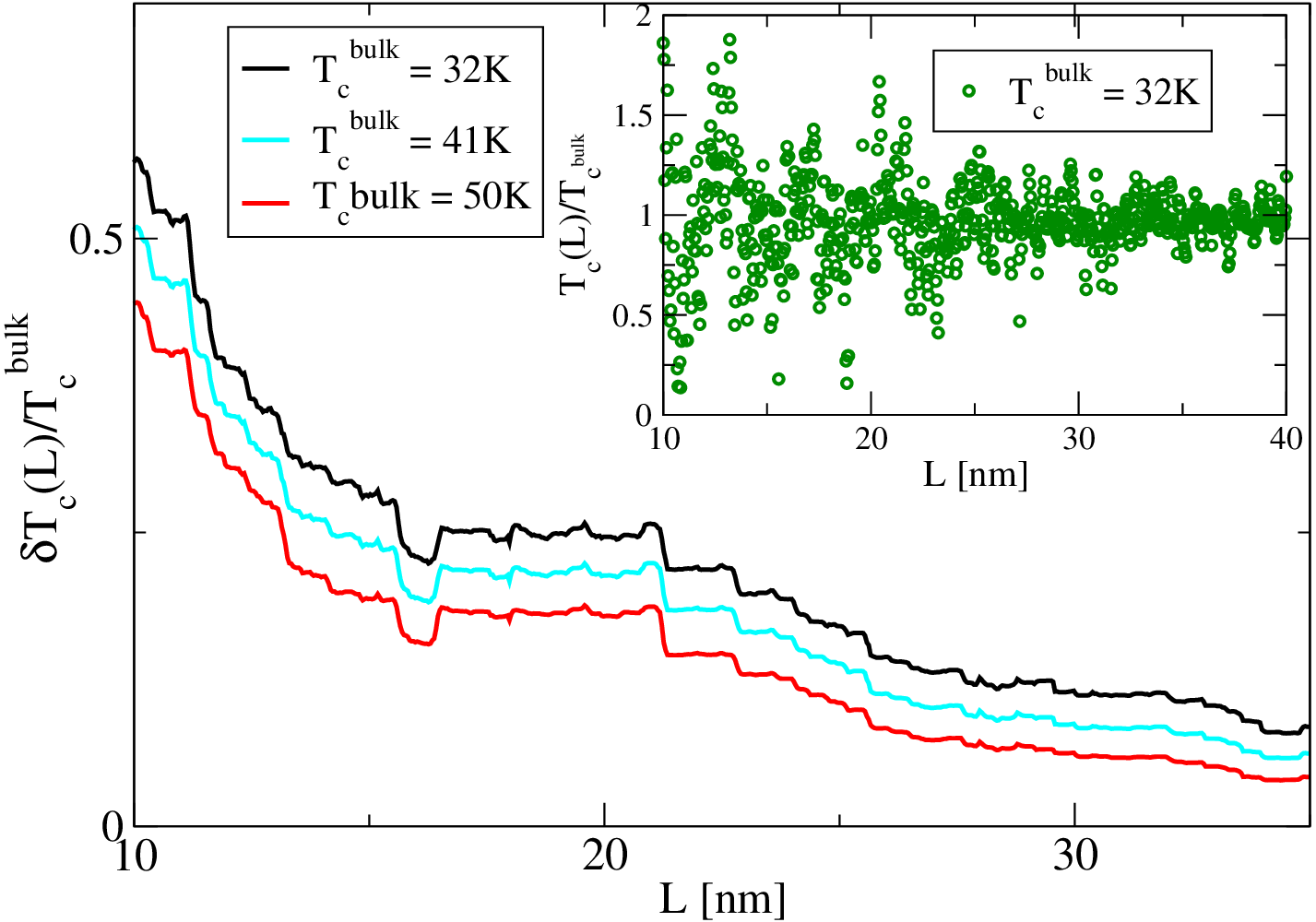}
\vspace{3mm}
\caption{$\delta T_c(L)/T_c^{\rm bulk}$ (typical deviation of $T_c(L)$ from the bulk limit $T_c^{\rm bulk} = T_c (L \to \infty) $ in a square film of side $L$ for different bulk $T_c$ and PBC. Upper: two band model (\ref{gap}). 
Lower: five band model \cite{graser}. Results of both models are qualitatively similar. Typical enhancement (suppression) of $T_c$ of more than $50 \%$ is 
observed for $L \sim 10$ nm.\\ 
Inset: Critical temperature $T_c(L)$ as a function of $L$ for $T_c^{\rm bulk} \approx 30$K.}
\label{fig1}
\end{figure}
\noindent{\it Models:}
We study FSE in two popular microscopic models of iron based superconductors: 
the minimal two band model of \cite{choi} and the five band model of \cite{graser}. It is important to investigate FSE in more than one model as, in these materials, there is not yet consensus about fundamental issues such as the mechanism for superconductivity (recent experiments \cite{neutron} suggest spin exchange interactions) or the 
leading pairing symmetry (see \cite{eremin,graser,dag,choi} for different proposals).\\
First we introduce the two band model of \cite{choi}:
the dispersion relation in the hole and electron band is given by, 
$\epsilon_{h} (k)=t_1 ^h (\cos k_x +\cos k_y) + t_2 ^h \cos k_x
\cos k_y + \epsilon^h$ and $\epsilon_{e} (k)=t_1 ^e (\cos k_x
+\cos k_y) + t_2 ^e \cos \frac{k_x}{2} \cos \frac{k_y}{2} +
\epsilon^e$. Following \cite{eremin,choi} we choose ($t_1, t_2, \epsilon$) to be
$(0.30,0.24,-0.6)$ for the hole band and $(1.14,0.74,1.70)$ for the electron
band. The bulk density of states is $N_h(0)=0.74/eV$ and $N_e(0)=0.285/eV$. Our results are robust to small changes of these parameters.  Superconductivity can be investigated by a mean field approach with two order parameters, $\Delta_h$ and $\Delta_e$, related by 
the following self consistent conditions:
\begin{eqnarray}
\Delta_h &=& -\sum_{k^{'} } V_{hh}\Delta_h \frac{\tanh (\frac{E_{h}(k')}{2 T})}{2E_{h}(k')} + V_{he} 
\Delta_e \frac{\tanh (\frac{E_{e}(k')}{2 T})}{2E_{e}(k')}  ,  \nonumber \\ \nonumber
\Delta_e &=& -\sum_{k^{'} } V_{eh} \Delta_h  \frac{\tanh (\frac{E_{h}(k')}{2 T})}{2E_{h}(k')} + V_{ee}
\Delta_e  \frac{\tanh (\frac{E_{e}(k')}{2 T})}{2E_{e}(k')}\\
\label{gap}
\end{eqnarray}
where, 
 $V_{he,he},V_{ee,hh}$ are the inter-band and intra-band coupling constants respectively and $E_{h,e} (k) =\sqrt{(\epsilon_{h,e}(k)-\mu)^2 + \Delta_{h,e}^2 (k)}$ are the energies of the quasiparticle excitations and $\mu$ is the chemical potential. The sums above are restricted to energies $\omega_{AFM} = 100$ meV around $\mu$. FSE are studied by simply solving (\ref{gap}) for a rectangular geometry. 
 We note that the different coupling constants are in principle momentum dependent $V_i = V_i(k,k')$. In \cite{choi} it was indeed employed a momentum dependent spin exchange interaction. It was however found that the order parameters were weakly dependent on momentum. Based on this result we have neglected any momentum dependence of the bulk interaction. 
In finite size systems the coupling constant gets a momentum/energy dependence \cite{blat,usprl,peeters} even if the bulk interaction is constant. Assuming a contact interaction, it is easy to show \cite{blat,peeters} that $V_i \propto I(k,k')$ where $I(k,k')$ are the so called  matrix elements $I(k,k') \propto L^2 \int dxdy \psi^2_k(x,y)\psi^2_{k'}(x,y)$ and $\psi_k$ are the eigenvectors of the one-body problem. In the case of chaotic grains \cite{usprl}, or rectangular grains with Dirichlet boundary conditions \cite{blat}, this correction induces a non-oscillatory enhancement of the energy gap in the nanoscale region (see inset Fig. \ref{fig2} (lower)). More specifically, the coupling constant effectively increases as $L$ decreases \cite{blat,peeters,usprl}. It is also clear from the definition of $I(k,k')$ that, in a rectangular system with periodic boundary conditions (PBC), the integral above is momentum independent. Due to this additional simplification we have decided to use PBC. We note that this is also a sensible choice to describe granular systems, where many grains are strongly coupled, or superconducting rings. Moreover results for other boundary conditions are qualitatively similar (see Fig. \ref{fig2}).\\  
We also investigate the five band model of \cite{graser} which provides a fully quantitative description of the band structure of an iron based superconductor within $2$ eV of the Fermi energy. For details of the model we refer to the appendix of \cite{graser}. Superconductivity, 
assumed to be of a $s^{\pm}$ type $\Delta(k) = \Delta(L)\cos(k_x)\cos(k_y)$, is 
studied in a BCS mean field fashion involving the five bands. 
FSE are investigated by solving exactly the mean field gap equations for a given geometry and $\omega_{AFM} = 80$ meV.\\
\begin{figure}
\includegraphics[height=0.5\columnwidth,clip,angle=0]{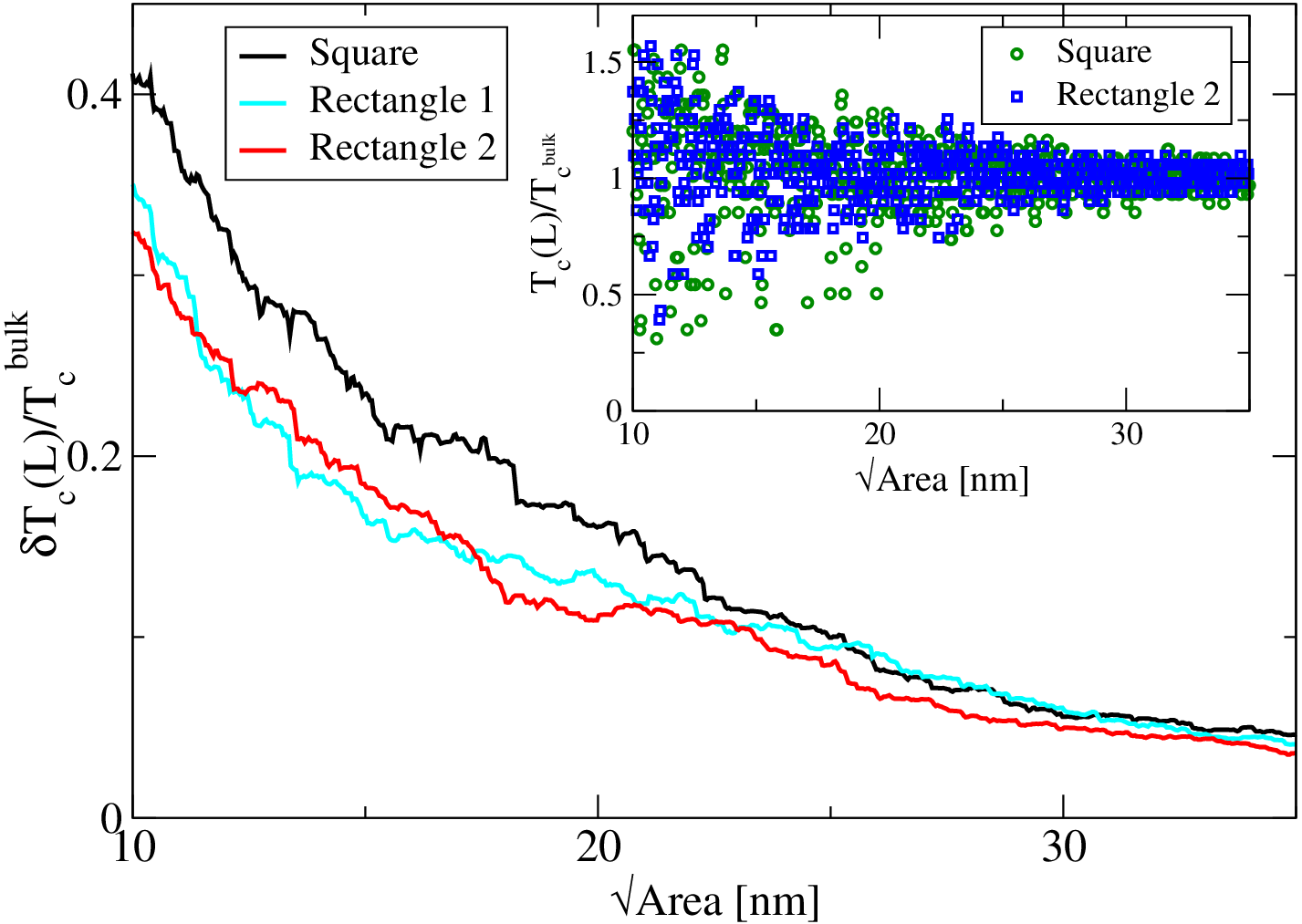}
\includegraphics[height=0.5\columnwidth,clip,angle=0]{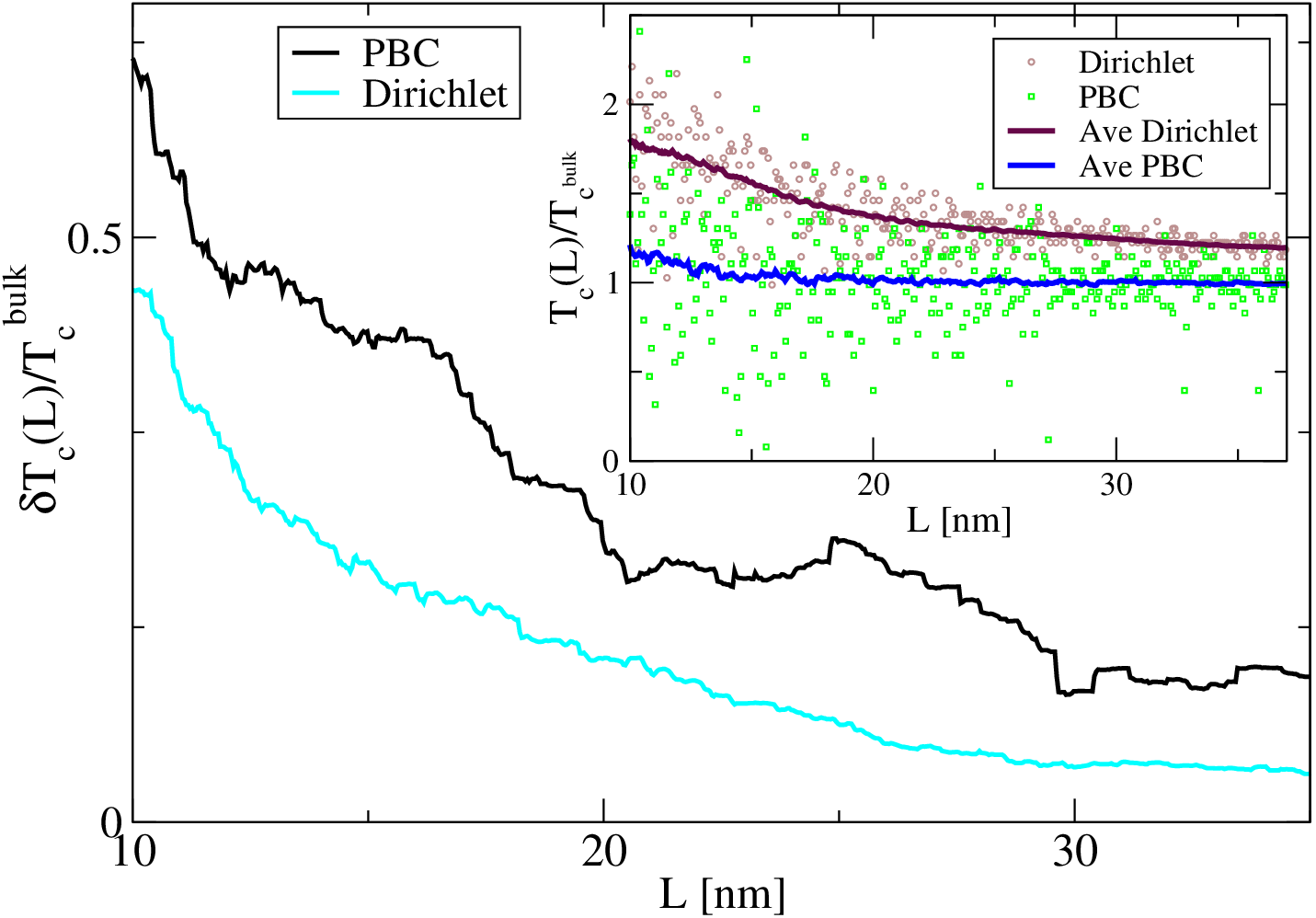}
\vspace{3mm}
\caption{Upper: $\delta T_c(L)/T_c^{\rm bulk}$ from (\ref{gap}) for a square of side $L$ and two rectangular films ($L_x = \alpha L, L_y = L/\alpha$) with $\alpha = 1.618034$ (rectangle $1$), $\alpha=1.2334$ (rectangle $2$) respectively and $T_c^{\rm bulk} \approx 32$ K. The intricate band structure and pairing pattern make the dependence on the shape weaker than in single band superconductors with parabolic spectrum. Inset: $T_c(L)$ for the square and rectangle. Lower: $\delta T_c(L)/T_c^{\rm bulk}$ in a square film, $T_c^{\rm bulk} \approx 32$K, for Dirichlet (including 
the effect of the matrix elements $I(k,k')$), and PBC. In the latter, {\it relative} deviations are stronger due to the additional level degeneracy but absolute deviations for small sizes can be still stronger for Dirichlet due to the effect of $I(k,k')$ (see text). Inset: Comparison between $T_c(L)$ for Dirichlet and PBC. The solid lines stand for the local $T_c(L)$ average in the 
same intervals ($\sim 3$nm) in which $\delta T_c(L)/T_c^{\rm bulk}$ is evaluated.}
\label{fig2}
\end{figure}
\noindent {\it Results:}
Our first task is to compute $T_c(L)$ (see inset Fig. \ref{fig1}) numerically for the two models above. Calculations are restricted to the range $L \in [10,40]$nm in order to keep the effect of thermal fluctuations small. For each $L$ the chemical potential $\mu$ was computed keeping constant the electron density. For the sake of simplicity we present results for a square film with PBC.
It will be shown that qualitatively similar results are also obtained for rectangular films and other boundary conditions (see Fig. \ref{fig2}). 
The value of the coupling constants are set to lead to typical values of $T_c^{\rm bulk} \in [30,50]$K in iron based superconductors.
FSE are rather insensitive to small changes of these values provided that $T_c^{\rm bulk}$ is still the same (see (\ref{tc})).   
For a more quantitative assessment of the typical maximum enhancement of $T_c$ we also plot
 $\delta T_c(L)/T_c^{\rm bulk}$, the typical deviation of $T_c(L)$ in units of $T_c^{\rm bulk}$. $\delta T_c(L)$ is obtained by evaluating the standard deviation of 
$T_c(L)$ in consecutive $\sim 3$ nm intervals where $T_c(L)$ is computed in steps of $0.04$nm. In Fig. \ref{fig1} we observe that: a) the typical deviations with respect to the $T_c^{\rm bulk}$ is quite large ($> 50\%$) in the $L \sim 10$ nm region, b) even for $L \approx 10$ nm and $T_c \approx 30$K thermal fluctuations are still small since $\gamma \sim 0.3 < 1$. We note that in our case $\delta$ is the geometrical mean (see (\ref{tc})) of the mean level spacing in each band, b) $T_c(L)$ is an oscillating function of $L$ with 
local maxima $T_c(L)$ much higher than in the $T_c^{\rm bulk}$ limit (see inset), c) FSE increase as $L$ or $T_c^{\rm bulk}$ decreases, d) results for both models are similar. 

In order to understand the origin of these features we study (\ref{gap}) in the limit $V_{ee} = V_{hh} = 0$ (this is a reasonable assumption as inter-band pair hopping is known to be the dominant mechanism) where an analytical estimation of $T_c (L)$ is feasible by using semiclassical techniques \cite{baduri,usprl}. For single band superconductors it was shown in \cite{usprl} that, in the region in which mean-field techniques are applicable, non-monotonic deviations from the bulk limit are well described by simply replacing sums by integrals in the gap equation, (\ref{gap}) in our case, and extracting from the integral a modified DOS $N_{\xi}(0)$. Technically this modified DOS is expressed as a sum over classical periodic orbits of the grain with lengths smaller than $\xi$. An explicit expression for a rectangular film can be found in \cite{baduri} and in the Appendix of the second reference of \cite{usprl}. In practical terms a similar result is obtained by simply smoothing the DOS $N(0) \sim \sum_i \delta(\epsilon - \epsilon_i)$ over an energy scale $\hbar v_F/\pi\xi$ where $v_F$ is the Fermi velocity. It is straightforward to show that a similar approach in our case leads to,
 \begin{equation}
T_c(L) \approx 1.136 \omega_{\rm AFM} \exp (-1/\lambda_{\rm eff})
\label{tc}
\end{equation}
where $\lambda_{\rm eff} = \lambda_{\rm bulk}\sqrt{\frac{N_{\xi_e}^e(0)N_{\xi_h}^h(0)}{N^e(0) N^h(0)}}$ with ${\lambda_{\rm bulk}}= \sqrt{V_{eh} N^{e}(0)V_{he} N^{h}(0)}$, and $\xi_{h(e)}$ stands for $\xi$ in the hole (electron) band. 

Several comments are in order: a) fluctuations in $T_c(L)$ are caused by finite size, not thermal, fluctuations of the spectral density around the Fermi energy. In systems with a parabolic spectrum it is well known that level degeneracy 
induced by geometrical symmetries \cite{baduri} of the film (cube, sphere...) increase these fluctuations, 
b) (small) changes in $\lambda_{\rm eff}$, due to FSE,  induce (exponentially) large changes in $T_c$. Therefore small changes in the number of levels around the Fermi energy can lead to large changes in $T_c$, c) FSE are strongly suppressed for $\xi/L \ll 1$ since, in this limit, the discrete nature of the spectrum is completely smoothed out and $\lambda_{\rm eff} \approx \lambda_{\rm bulk}$. Since $\xi$ decreases as $\lambda_{\rm bulk}$ increases, deviations from the bulk limit for a given size $L$ decrease as $T_c^{\rm bulk}$ increases. This is what is observed in Fig. \ref{fig1}.\\
The rest of predictions are now tested numerically for the two-band model (\ref{gap}). More specifically we study the dependence of FSE on boundary conditions and the shape of the grain.
In Fig. \ref{fig2} (upper) we compare the typical deviation $\delta T_c(L)/T_c^{\rm bulk}$ for two different rectangles and a square of the same total area. Deviations from the bulk limit are slightly larger in the square. This is due to the existence of a
symmetry $k_x \to k_y$ in the spectrum of the square which is not present in the rectangle. As was mentioned previously, level degeneracy in the spectrum leads to stronger fluctuation in the number of levels around the Fermi energy and, according to (\ref{tc}), to larger deviations of $T_c(L)$ from $T_c^{\rm bulk}$. We note however that, the difference between the two geometries is much smaller than in single band superconductors with parabolic spectrum. The reason for that is that the level degeneracy of a parabolic spectrum is much larger than the one ($k_x \to k_y$ symmetry) present in the models we study.
\begin{figure}
\includegraphics[height=0.6\columnwidth,clip,angle=0]{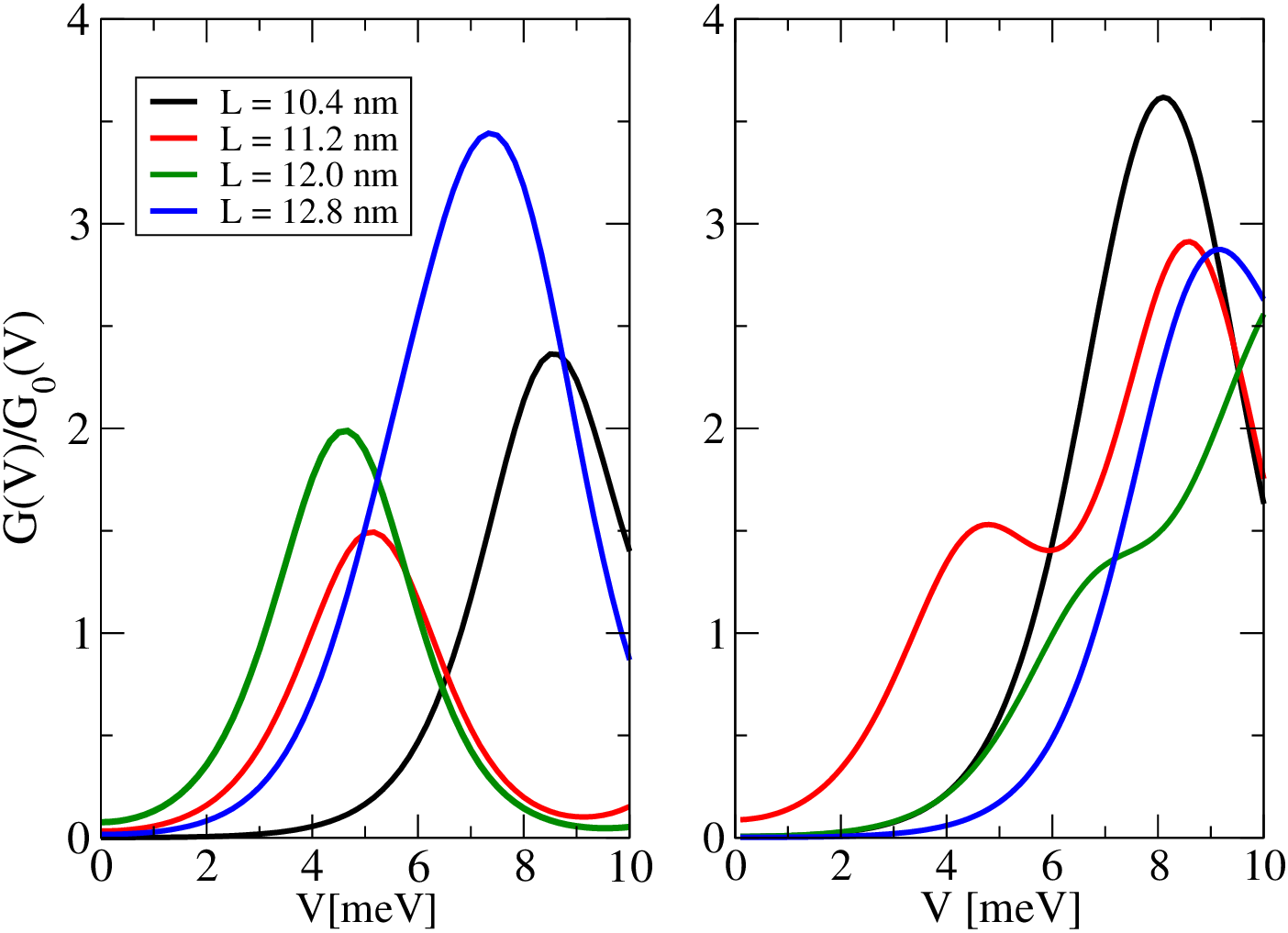}
\vspace{3mm}
\caption{
Differential conductance $G(V)$ (\ref{gv}) as a function of the bias voltage $V$ for $T \approx T_c/3$, $T_c \approx 31$ K and different sizes.    $G(V)$ is rescaled by its value, $G_0(V)$, at $T \approx 2 T_c$. Left: Two band model (\ref{gap}). The peak in $G(V)$ coincides with $min(|\Delta_e|,|\Delta_h|)$ in (\ref{gap}). Right: Five band model \cite{graser}. The peak closer to $V = 0$ also corresponds with the value of the energy gap. The additional structure is a consequence of the combined effect of the band structure and level degeneracy. The substantial differences observed for small changes in $L$ suggest that it is experimentally feasible to detect FSE by scanning tunnelling microscope techniques (STM).}
\label{fig3}  
\end{figure} 
 In Fig. \ref{fig2} (lower) we compare $\delta T_c(L)/T_c^{\rm bulk}$ for a square with PBC and Dirichlet boundary conditions including the effect of the matrix elements $I(k,k')$. 
 It is observed that, for Dirichlet boundary conditions, {\it relative} fluctuations are smaller. This is due to the additional level degeneracy for PBC where eigenstates with opposite momentum have the same eigenvalue. By relative we mean that, as in the other cases, the standard variance is computed with respect to the local average $T_c$, not the $T_c^{\rm bulk}$. As can be observed in the inset of Fig. \ref{fig2} (upper), 
 the overall effect of $I(k,k')$ is to enhance significantly \cite{blat,usprl} the average $T_c(L)$. As a result, the absolute deviation from the bulk limit for Dirichlet boundary conditions is of the order or even larger than for PBC.
In summary, in the range of $T_c^{\rm bulk} \in [20,50]K$ typical of iron-pnictides materials, substantial deviations from the bulk limit are observed for different models, shapes and boundary conditions for sizes $ L \sim 10$nm for which thermal fluctuations are not yet important.
 
Finally we discuss to what extent is possible to observe experimentally FSE. 
In metallic nanograins the normalized differential conductance, 
\begin{equation}
\label{gv}
G(V)  =  \frac{1}{4Tk_B}\int_{-\infty}^{\infty}d\omega\, N_{s}(\omega)\left[\frac{1}{\cosh^{2}\left(\frac{\omega+V}{2k_{B}T}\right)}\right]
\end{equation}
(where $N_{s}(\omega)$ stands for the superconducting DOS) is one of the most popular experimental 
observables as it can be measured by STM techniques. 
In Fig. \ref{fig3} we depict $G(V)$ as a function of the bias voltage $V$ in the two models above for different, but similar sizes. It is clearly observed that, in the nanoscale region, $G(V)$ is highly sensitive to both FSE and the model used. Small changes in the system size induce strong modifications in $G(V)$. This is another indication that it is in principle possible to observe and study experimentally FSE by STM techniques. More precisely we believe that, 
in order to compare our results with experiments, the following conditions must be met: a) Fermi liquid theory and mean-field techniques must be  applicable, b) it is technically feasible to manufacture single grains or granular materials of size $L \sim 10$ nm for which thermal fluctuations are negligible but FSE are still relevant, c) the theoretical 
prediction for $\xi$ cannot be very different from the experimental one. Otherwise the experimental enhancement of FSE, which  is 
 controlled by the ratio $\xi/L$, will not occur in the range of sizes we are proposing. We have checked for the model of \cite{graser} that the averaged $\xi$ over the Fermi surface is consistent with the experimental one. 
 
In summary, we have investigated FSE in superconducting iron-pnictides nanostructures. Within a mean field approach we have identified a region $L \sim 10$ nm in which FSE can enhance $T_c$ substantially and thermal fluctuations, which break long range order, are not important. Qualitatively similar results are obtained for different boundary conditions, system geometries and band structure. Experimental observation of these effects is in principle feasible as observables such as the differential conductance are very sensitive to FSE.\\

\noindent We acknowledge partial support from Projects
PTDC/FIS/101126/2008, PTDC/FIS/111348/2009. AMG acknowledges partial financial support from a Marie Curie International Reintegration Grant PIRG07-GA-2010-268172.

\end{document}